\documentclass[12pt]{spieman}  % 12pt font required by SPIE;
\usepackage{amsmath,amsfonts,amssymb}
\usepackage{graphicx}
\usepackage{setspace}
\usepackage{tocloft}

\title{Holographic particle localization under multiple scattering}

\author[a]{Waleed Tahir}
\author[b,c]{Ulugbek S. Kamilov}
\author[a,*]{Lei Tian}
\affil[a]{Department of Electrical and Computer Engineering, Boston University, Boston, MA 02215, USA.}
\affil[b]{Department of Electrical and Systems Engineering, Washington University in St.~Louis, MO 63130, USA.}
\affil[c]{Department of Computer Science and Engineering, Washington University in St.~Louis, MO 63130, USA.}

\cftpagenumbersoff{figure}
\cftpagenumbersoff{table} 

\usepackage{mathrsfs,cleveref}

\newcommand{\mb}{\mathbf}

\newcommand{\tb}[1]{\textbf{#1}}
\newcommand{\tn}[1]{\text{#1}}
\newcommand{\conj}[1]{\overline{#1}}

\begin{document} 
\maketitle

\begin{abstract}
	We introduce a new computational framework that incorporates multiple scattering for large-scale 3D particle localization using single-shot in-line holography. 
	Traditional holographic techniques rely on single-scattering models that become inaccurate under high particle densities and large refractive index contrasts. 
	Existing multiple scattering solvers become computationally prohibitive for large-scale problems, which comprise of millions of voxels within the scattering volume.
	Our approach overcomes the computational bottleneck by slice-wise computation of multiple scattering under an efficient recursive framework.
	In the forward model, each recursion estimates the next higher-order multiple scattered field among the object slices.
	In the inverse model, each order of scattering is recursively estimated by a novel nonlinear optimization procedure.  
	This nonlinear inverse model is further supplemented by a  sparsity promoting procedure that is particularly effective in localizing 3D distributed particles. 
	We show that our multiple scattering model leads to significant improvement in the quality of 3D localization compared to traditional methods based on single scattering approximation.
	Our experiments demonstrate robust inverse multiple scattering, allowing reconstruction of 100 million voxels from a single 1-megapixel hologram with a sparsity prior.
	The performance bound of our approach is quantified in simulation and validated experimentally.
	Our work promises utilization of multiple scattering for versatile large-scale applications.
\end{abstract}

% Include a list of up to six keywords after the abstract

% Include email contact information for corresponding author
{\noindent \footnotesize\textbf{*}  \linkable{leitian@bu.edu} }

\begin{spacing}{2}   % use double spacing for rest of manuscript

\section{Introduction}
	Three dimensional (3D) particle-localization using in-line holography is fundamental to many applications, such as biological sample characterization~\cite{moon2009automated,su2012high}, flow cytometry~\cite{Cheong:09,garcia2006digital}, fluid mechanics~\cite{sheng2006digital,Tian.etal2010}, and optical measurement~\cite{picart2007tracking,schnars2002digital,soulez2007inverse}.
	Reconstructing dense samples, however, still remains  challenging~\cite{Chen.etal2015a}.  
	Standard back-propagation method (BPM) can only handle low particle density~\cite{Chen.etal2015a}.  
	Compressive holography based on the first Born approximation significantly improves upon BPM by imposing sparsity constraints~\cite{Brady.etal2009,rivenson2010compressive}. 
However, it suffers from an underlying single scattering assumption, which becomes invalid at high particle densities where multiple scattering effects become significant.  
In this work, we propose, for the first time to our knowledge, a framework that accounts for multiple scattering in in-line digital holography, and enables accurate 3D particle localization at high density in a computationally efficient fashion.

Multiple scattering induces a {\it nonlinear} relation between the permittivity contrast and the scattered field, making it difficult to invert~\cite{Born.Wolf2003}. 
	Many algorithms have been proposed to solve the inverse multiple scattering problem and demonstrated improved performance over single-scattering methods, such as iterative Born series~\cite{Chew.Wang1990,chen1998experimental,Kamilov.etal2016a,tsihrintzis2000higher,pierri1999inverse}, contrast source inversion~\cite{vandenBerg.Kleinman1997, Bevacquad.etal2017,van1999extended,van2001contrast}, modified gradient~\cite{Kleinman.vandenBerg1992,Kleinman.vandenBerg1993}, Series Expansion with Accelerated Gradient Descent on Lippmann-Schwinger Equation (SEAGLE)~\cite{liu2018seagle,ma2017accelerated}, and hybrid methods \cite{pham2018versatile,lim2017beyond,Simonetti2006,Mudry.etal2012}. 
	However, computational challenges restrict them to be demonstrated only for small-scale problems. 
	This is because modeling multiple scattering necessitates computing the {\it internal} scattered field within the object volume.	Furthermore, the effectiveness of existing multiple scattering methods has been demonstrated only under  multi-shot tomography. 
	While multiple measurements do alleviate the ill-posedness of the inverse problem, they also increase acquisition time and system complexity.
	%Multi-shot also poses a challenge when imaging dynamic objects. 
	Therefore, it is  of interest to investigate if one can exploit multiple scattering using only a single-shot measurement. 
	In this work, we demonstrate successful inverse multiple scattering for large-scale problems and reconstruct 100~million voxels from a single 1-megapixel in-line hologram. 
	We show that even under such highly ill-posed conditions, inversion of multiple scattering is still possible and can be used to improve results compared to single scattering techniques.
	
	To calculate multiple scattering, we build our model based on the Born series expansion~\cite{Born.Wolf2003}. 
	To make it computationally efficient, we take a multislice approximation by discretizing the 3D object volume into a series of 2D thin axial slices. 
	At each slice, each object voxel takes a uniform refractive index value.
	Between neighboring slices, the uniform background medium is assumed. 
	By adjusting the voxel size and inter-slice distance, our model allows to flexibly trade off computational complexity for model accuracy. 
	At the limit when the voxel size equals the inter-slice distance, our discretization reduces to the existing approaches in~\cite{Chew.Wang1990,chen1998experimental,Kamilov.etal2016a}. 
	Our computational structure closely resembles the multislice model (\emph{i.e.}~beam propagation method)~\cite{Tian.Waller2015,kamilov2016optical,kamilov2015learning}.
	However, our model has the benefit of computing {\it both forward and backward} scattering, whereas the latter only accounts for {\it forward} multiple scattering.
	
	To compute multiple scattering, we introduce a novel 3D-to-3D operator to efficiently evaluate the {\it internal} scattered fields within the volume. The computational framework discretizes the 3D object as a set of 2D slices, and multiple scattering is modeled as recursive propagation among them.
	Starting from the initial field, each subsequent recursion estimates the next higher-order scattering term within the object volume. 
	This process can be carried out up-to an arbitrary order until the field converges to a steady-state. 
	To evaluate the convergence, we adapt a metric derived from the residual error of the internal field~\cite{vandenBerg.Kleinman1997,liu2018seagle}.
	Next, we devise a 3D-to-2D operator that computes the {\it external} scattered fields by propagating the multiply scattered internal 3D field to the 2D sensor plane.
	Finally, the intensity measured by the hologram is the interference between the scattered  and the unscattered fields. 
	This further complicates the model by introducing the ``twin-image'' problem~\cite{guizar2012understanding}.
	If only single scattering is considered, our model reduces to linear compressive holography.~\cite{Brady.etal2009}. 
	As a result of multiple scattering computation, the hologram encodes information about the high-angle scattering within the volume, which is otherwise ignored in single scattering based methods.
	We show that this extra information leads to better recovery of the scatterers, in particular at larger depths.

To solve the inverse scattering problem, we derive an optimization procedure that iteratively minimizes the data-fidelity term measuring the difference between the estimated and measured holograms, and imposes a sparsity-promoting regularization on the object. 
The overall structure of the algorithm follows the proximal-gradient method~\cite{Beck.Teboulle2009b}. 
The key ingredient is the gradient computation of the data-fidelity term.
Conveniently, our recursive forward model leads to a similarly structured recursive gradient computation.
Further exploiting the convolution structure in the scattering operators, the algorithm is implemented using efficient FFT-based computations.

	%This prior is particularly effective for discrete scatterers suspended in a homogeneous medium. 
	Distinct from prior Born-series based models~\cite{Chew.Wang1990,chen1998experimental,Kamilov.etal2016a,tsihrintzis2000higher,pierri1999inverse}, we do not directly measure the full complex field.  
	The effect of sparsity regularization to the twin-image artifact has been studied using single scattering~\cite{Brady.etal2009,zhang2018twin}, and 2D multiple scattering models~\cite{pham2018versatile}.
	Here, we show that the sparsity is also effective in suppressing twin-image artifacts under 3D multiple scattering models.
	
An important feature of our multislice-based framework is that the 3D object can be flexibly estimated with any desired number of axial slices, as set by the targeted resolution. 
In particular, we show that it is possible to use much fewer axial slices to achieve high  localization accuracy while still exploiting  the extra information contained in the multiple scattering.  
This allows us to handle much larger scale problems with reduced computational cost as compared to existing techniques that are often limited by fine sampling requirements.

Single scattering based methods tend to under-estimate the refractive index contrast.
This under-estimation can be mitigated by incorporating multiple scattering~\cite{tsihrintzis2000higher,liu2018seagle,Ling2018,lim2018learning}.
We show this effect using our multislice-based approach in single-shot in-line holography, and demonstrate improved particle localization and axial resolution under multiple scattering.

Next, we demonstrate the localization accuracy of our method by imaging 3D distributed particles in water at various densities in both simulation and experiment. 
To facilitate quantitative comparison of different methods, we use a classification framework and use the receiver operating characteristic (ROC) curve to determine each method's best performance.
At low particle density, our multiple scattering model converges to the single scattering solution as expected, since the information is dominated by the first order scattering. 
At high particle density, our model largely improves the accuracy since multiple-scattering becomes more significant. 
We observe that the localization accuracy is highly depth-dependent. 
Following the classification framework, we use the dice coefficient~\cite{Rokach2005} to quantify the localization result slice-by-slice.
We show that our multiple scattering model provides greater improvement at larger depths.  

%Our technique paves the way for many versatile large-scale applications for multiple scattering.

	% intro figure
	\begin{figure}[t]
		\centering
		\includegraphics[width=0.7\linewidth]{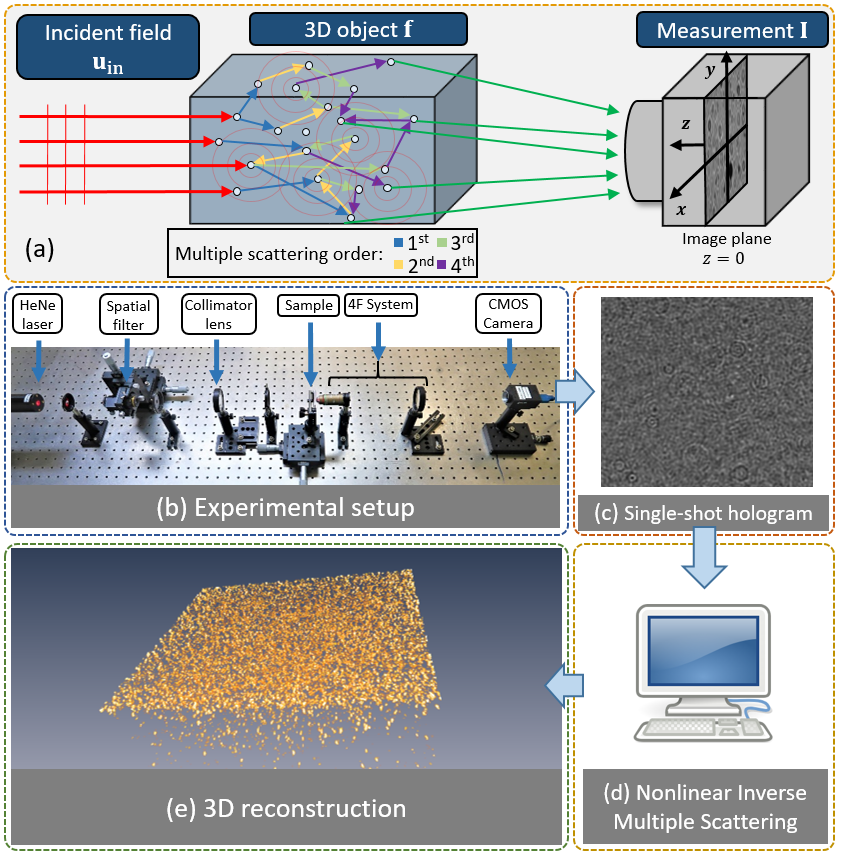}
		\caption{In-line holography with multiple scattering. 
		(a)~A plane-wave is incident on a 3D object containing distributed scatterers. The field undergoes multiple scattering within the volume, then propagates to the image plane. A hologram is recorded, which is then used to estimate the unknown scatterers' distribution. 
		(b) An inline holography setup is used that consists of a collimated laser for illumination and a 4F system for magnification. 
		(c) The raw data is a single hologram.
		(d) The reconstruction implements a nonlinear inverse multiple scattering algorithm.~\cite{gnome} 
		(e) The output estimates the 3D distribution of the scatterers.}
		\label{fig:setup}
	\end{figure}

\section{Theory and method}
	\label{sec:method}
	
	\subsection{Forward model}  \label{Forward model}
	Consider the imaging geometry in Fig.~\ref{fig:setup}(a). 
	An in-line hologram $I_{\tn{m}}$ at the measurement plane can be written as 
	\begin{equation}
	  I_{\tn{m}}(\mb{x}_\tn{0}) = |u_\tn{in}(\mb{x}_\tn{0},0)+E(\mb{x}_\tn{0},0)|^2 = 
	  |u_\tn{in}|^2 + 2u_\tn{in}\tn{Re}\{E\}+
	  |E|^2,
	\end{equation}
	where $E$ is the scattered field on the measurement plane, $u_\tn{in}$ is the incident plane wave and is assumed to be real on the hologram plane ($z_\tn{0}=0$) without loss of generality, $\mb{x}_\tn{0}=(x_\tn{0},y_\tn{0})$ represents the transverse spatial coordinates on the hologram plane. 
	The self-interference term of the scattered field $|E|^2$ is ignored; the validity of this assumption is discussed in Sec.~\ref{sec:largescalesim}. 
	From Eq.~(\ref{hologram}), the scattered field and its ``twin-image'' contribution is related to the measured hologram after background removal by 
	\begin{equation}
	    \tn{Re}\{E(\mb{x}_\tn{0},0)\} = \frac{[E(\mb{x}_\tn{0},0)+E^*(\mb{x}_\tn{0},0)]}{2}= \frac{I_{\tn{m}}(\mb{x}_\tn{0})-|u_\tn{in}|^2}{2u_\tn{in}}.
	    \label{hologram}
	\end{equation}

	\begin{figure}[t]
		\centering
		\includegraphics[width=0.65\linewidth]{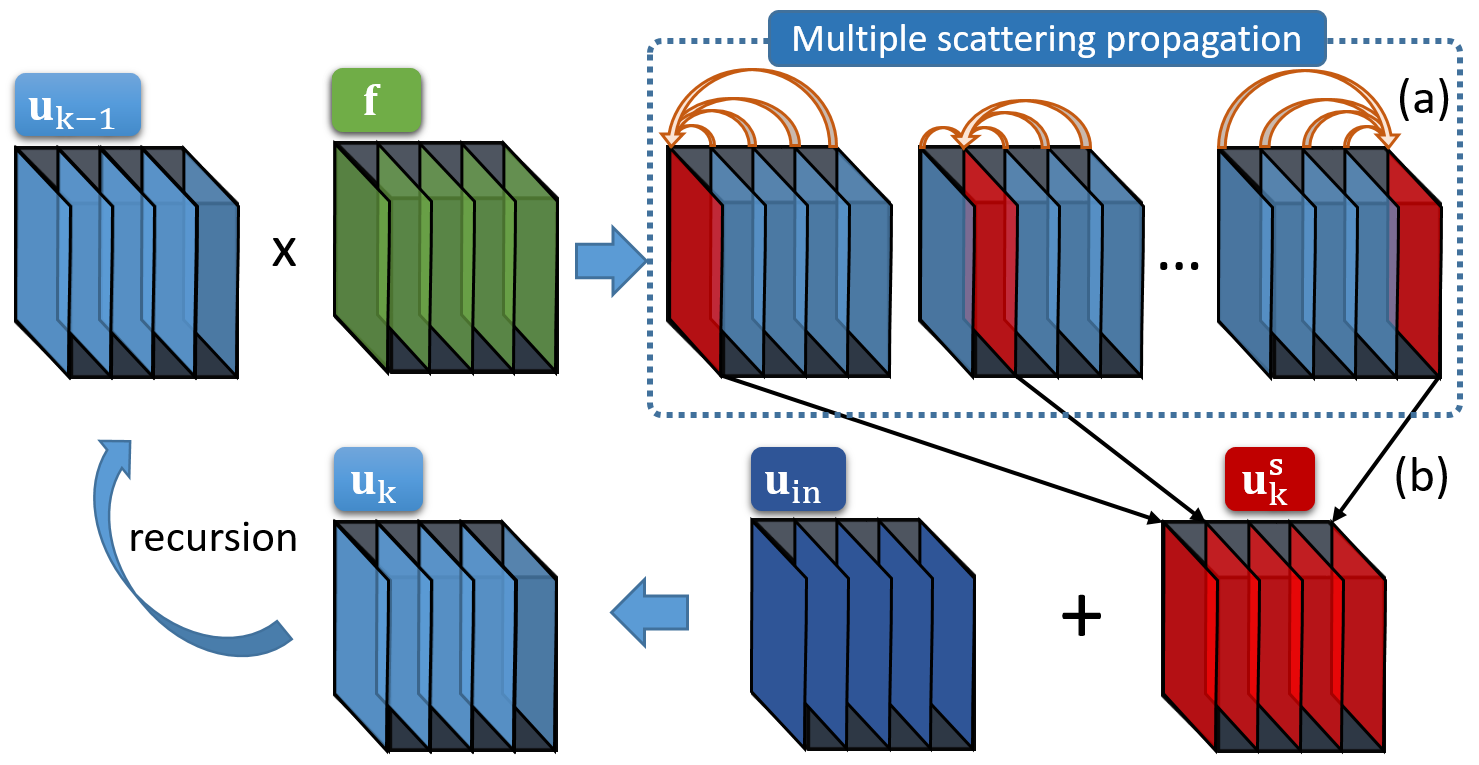}
		\caption{Illustration of the 3D internal scattered field operator $\mb{G}$ in Eq.~(\ref{e04-fmd}). 
		(a) Each object slice  $\mb{f}$ is first voxel-wise multiplied by the lower order scattered field $\mb{u}_\tn{k-1}$; it is then  propagated to every other slice within the volume. 
		(b) This computed scattered-field $\mb{u}^\tn{s}_\tn{k}$ is  added to the incident-field $\mb{u}_\tn{in}$ to obtain the next higher-order Born-field $\mb{u}_\tn{k}$. 
		This process is recursively applied to compute the multiply scattered field within the volume.}
		\label{fig:G}
	\end{figure}

	The background-removed hologram thus represents the real component of the scattered field at the measurement plane, and is given as  $I_{\tn{br}}(\mb{x}_\tn{0})
	  =
	  (I_{\tn{m}}(\mb{x}_\tn{0})-|u_\tn{in}|^2)/2u_\tn{in}$.
	To model the hologram resulting from multiple scattering up-to the $\tn{K}^{\tn{th}}$~order, we apply the framework of Born series expansion~\cite{Born.Wolf2003,Kamilov.etal2016a}, which gives two coupled equations:
	\begin{align}
	& E(\mb{x}_\tn{0},0) = \iiint_{\Omega}{f(\mb{x}',z') u_\tn{K}(\mb{x}',z') h(\mb{x}_\tn{0}-\mb{x}',-z') d^2\mb{x}'dz'},
	\label{e01-iba1} \\
	& u_k(\mb{x},z)  = u_\tn{in}(\mb{x},z) 
	 + \iiint_{\Omega}{f(\mb{x}',z') u_{k-1}(\mb{x}',z') h(\mb{x}-\mb{x}',z-z') d^2\mb{x}'dz'},\label{e02-iba2}
	\end{align}
	where $h$ is the 3D Green's function and $u_\tn{K}(\mb{x'},z')$ is the $\tn{K}^{\tn{th}}$~order multiply scattered field within the volume $\Omega$. 
	This mathematical model is based the scalar Helmholtz equation~\cite{Born.Wolf2003}, and polarization effects are neglected.
	This internal field is computed recursively within the support $\Omega$ using the Born series [Eq.~\eqref{e02-iba2}].
The permittivity contrast $f$ is related to the refractive index by $f(\mb{x}',z')=\frac{k^2}{4\pi}({n^2(\mb{x}',z') - n_\tn{med}^2})$~\cite{Born.Wolf2003}, where $n_\tn{med}$ is the index of the homogeneous background medium, and $k$ is the wave number in free space.
For simplicity, $f$ is assumed to be real valued and absorption effects are ignored.
	We note that the spatial coordinates associated with the object are within the 3D support: $(\mb{x},z)\in \Omega$, $(\mb{x}',z')\in \Omega$; whereas the hologram is measured outside the support: $(\mb{x}_\tn{0},z_\tn{0})\notin \Omega$. To compute higher order scattering, the initial condition is $u_0(\mb{x},z) = 0$, and $k = 1,2,...,\tn{K}$ indexes the scattering order. 
	When $\tn{K}=1$, $u_1(\mb{x},z) = u_\tn{in}(\mb{x},z)$ is the incident field [from Eq.~\eqref{e02-iba2}], and 	Eq.~\eqref{e01-iba1} reduces to the first-Born approximation that {\it linearly} relates the object to the {\it singly scattered} field. 
	When $\tn{K}=2$, this relation becomes {\it nonlinear} and the second order {\it multiple scattering} is taken into account via modeling of the additional interaction between the object and the field within the volume. 
	For larger K, the approximation becomes more accurate by accounting for K-order multiple scattering.
	
	Equations~(\ref{e01-iba1}) and (\ref{e02-iba2}) can be discretized to get the following {\it recursive forward model}
	\begin{align}
	&\mb{E} = \mb{H}~\tn{diag}(\mb{u}_\tn{K})~\mb{f},\label{e03-fmd}\\
	&\mb{u}_k = \mb{u}_\tn{in} + \mb{G}~\tn{diag}(\mb{u}_{k-1})~\mb{f},\label{e04-fmd}
	\end{align}
	for $k=1,2,...,\tn{K}$. Here, $\mb{f}$ and $\mb{u}_k$ have dimension $(N_x N_y N_z) \times 1$, and $\mb{E}$ is $(N_x N_y) \times 1$. $N_x$, $N_y$, and $N_z$ are the number of pixels along the $x$, $y$, and $z$ within $\Omega$, respectively. $\tn{diag}(\mb{u})$ represents a diagonal matrix with the vector $\mb{u}$ on the main diagonal.
	% H operator
	% dim of Q0

	We consider the 3D volume to be a set of discrete 2D slices along the longitudinal axis. $\mb{H}$ and $\mb{G}$ are the scattering operators which represent propagation among the object slices and to the hologram plane. 
	$\mb{H}$ propagates the field from the object support to the hologram plane. 
	$\mb{H} = \mb{K}^H \mb{Q}_0 \mb{B}$, where $\mb{B} = \mathrm{bldiag}(\mb{K},..,\mb{K})$ is a block-diagonal matrix. $\mb{K}$ and $\mb{K}^H$ are the 2D DFT and the inverse 2D DFT matrices respectively, each with dimension $(N_x N_y) \times (N_x N_y)$. 
	$\mb{Q}_0 = [\mb{L}_{0-1}~\mb{L}_{0-2}~...~\mb{L}_{0-N_z}]$, where $\mb{L}_{z'-z}$ is a diagonal matrix representing the discrete transfer function that performs propagation between two slices, from the slice $z$ to $z'$. This treatment of the $\mb{H}$ operator is similar to that in~\cite{Brady.etal2009}.
	% G operator
	$\mb{G}$ is {\it the multiple scattering operator} that performs propagation from the object volume to within itself (Fig.~\ref{fig:G}). 
	$\mb{G} = \mb{B}^H \mb{R} \mb{B}$, where $\mb{R} = [\mb{Q}_1, \mb{Q}_2, ..., \mb{Q}_{N_z}]$ and $\mb{Q}_m = [\mb{L}_{m-1}~\mb{L}_{m-2}~...~\mb{L}_{m-N_z}]$. 
	$\mb{R}$ has the dimension of  $(N_x N_y N_z) \times (N_x N_y N_z)$, and contains transfer functions that propagate the field from each slice to every slice within the support.
	There are two methods of computing the elements in the  transfer function $\mb{L}_{z'-z}$ having dimension $(N_x N_y) \times (N_x N_y)$, the direct and the angular spectrum methods \cite{Mas1999}. 
	We use the direct  method, in which the Green's function $h(\mb{r})=\exp(\dot{\iota}k |\mb{r}|)\big/|\mb{r}|$ is sampled in the spatial domain, followed by slice-wise 2D FFT, where $\mb{r}=(x,y,z)$. 
	
	An important numerical treatment to $h(\mb{r})$ is around the singularity at $|r|=0$. 
	We adapt the technique in~\cite{Yaghjian1980,VanBladel1961} and consider a spherical exclusion zone around $|r|=0$ of radius $a$, inside which the Green's function is assumed to take a constant value. 
	Effectively, this assigns an ``averaged'' Green's function value around the singular region. 
	Empirically, we found that at low refractive index, the choice of $a$ does not significantly affect the result as long as the center voxel (at $|r|=0$) is excluded. 
	For high refractive index, $a$ highly affects the convergence of the forward model.  
	We set $a$ to match the largest expected radius of the particles.
	This means that the strong multiple scattering inside each individual particle cannot be reliably modeled at high contrast; hence it is ignored.
	Only particle-particle interactions are modeled. 
	Correspondingly, during the inversion, $a$ sets the largest particle size that can be recovered by our model for high index particles.

	\subsection{Inverse problem}
	To estimate the object $\mb{f}$ from the holographic measurement, we need to solve Eqs.~(\ref{e03-fmd},\ref{e04-fmd}). 
	Unlike traditional digital holography, this problem is nonlinear when $\tn{K}>1$. 
	We devise an inverse scattering algorithm~\cite{Fannjiang2013,Fannjiang2015} that minimizes a cost function $\mathcal{C}(\mb{f})$ to compute the estimated object $\hat{\mb{f}}$ as follows
	\begin{equation}
	\hat{\mb{f}} =
	\underset{\mb{f}\epsilon\mathcal{F}}{\tn{argmin}}
	\{\mathcal{C}(\mb{f}) \} = 
	\underset{\mb{f}\epsilon\mathcal{F}}{\tn{argmin}}\{\mathcal{D}(\mb{f}) + \tau \|\mb{f}\|_{\textsf{\tiny TV}}\},\label{e05-inv}
	\end{equation}
	where 
	$\mathcal{D}(\mb{f}) = \frac{1}{2}\|\mb{I}_\tn{br}-\tn{Re}\{ \mb{E}_\tn{est}\}\|^2$ 
	is the data fidelity term in which $\mb{I}_\tn{br}$ is the real valued measured hologram after background removal, and $\mb{E}_\tn{est}$ is the complex-valued scattered field estimate from our model [Eq.~\eqref{e03-fmd}]. $\tn{Re}\{ \mb{E}_\tn{est}\}$ provides an estimate for $\mb{I}_\tn{br}$ [Eq.~\eqref{hologram}], $\|\cdot\|$ represents the L2-norm, $\mathcal{F}$ is the convex set that constrains the object to be nonnegative, and 
	$\tau$ is the regularization parameter that is empirically tuned.
	$\|\mb{f}\|_{\textsf{\tiny TV}}$ imposes a penalty on the total variation (TV) of the object, and is defined as
	\begin{equation}
	\|\mb{f}\|_{\textsf{\tiny TV}}
	\overset{\Delta}{=}
	\sum_{n=1}^{N} \|[\mb{D}\mb{f}]_n\|_{l_2}
	=
	\sum_{n=1}^{N}
	\sqrt{|[\mb{D}_x\mb{f}]_n|^2
	+
	|[\mb{D}_y\mb{f}]_n|^2
	+
	|[\mb{D}_z\mb{f}]_n|^2},
	\end{equation}
	where $\mb{D} : \mathbb{R}^N \rightarrow \mathbb{R}^{N\times3}$ is the discrete gradient operator with matrices $\mb{D}_x$, $\mb{D}_y$, and $\mb{D}_z$ denoting the finite difference operations along $x$-, $y$-, and $z$-directions, respectively. 
	 
	The minimization in Eq.~\eqref{e05-inv} is implemented via the proximal-gradient method~\cite{beck2009fast}, in which the $t^{\tn{th}}$ iteration is written as
	\begin{equation}
	\mb{f}^\tn{t} \leftarrow \tn{prox}_{\tau \textsf{\tiny{TV}}}\left( \mb{f}^{\tn{t}-1} - \alpha \frac{\partial \mathcal{D}(\mb{f})}{\partial \mb{f}} \right),\label{e06-prox}
	\end{equation}
	where $\tn{prox}_{\tau \textsf{\tiny{TV}}}\big(\mb{g}) \overset{\Delta}{=} \underset{\mb{f}\epsilon\mathcal{F}}{\tn{argmin}} \big\{\frac{1}{2}\|\mb{f} - \mb{g}\|^2 + \tau \|\mb{f}\|_\textsf{\tiny{TV}}\big\}$ is the proximal operator for TV minimization~\cite{kamilov2017parallel}, and $\alpha$ is the step size set via backtracking line search~\cite{Andrei2006}. 
	%Each iteration in the reconstruction is thus a dual-step process with a gradient descent step followed by TV-denoising. 
	The initialization is $\mb{f}^0 = \mb{0}$.
	
	Similar to the forward model, the gradient computation is also a K-order recursion.
	\begin{align}
	&\frac{\partial \mathcal{D}(\mb{f})}{\partial \mb{f}}
	= 
	\tn{Re}\bigg\{
	\tn{diag}(\conj{\tb{u}}_\tn{K})\tb{H}^H\mb{r}+
	\Big[
	\frac{\partial \tb{u}_\tn{K}}{\partial \mb{f}}
	\Big]^H
	\tn{diag}(\mb{f})
	\tb{H}^H\mb{r}
	\bigg\},\label{e07-grad}\\
	&\Big[
	\frac{\partial \tb{u}_{k}}{\partial \mb{f}}
	\Big]^H
	\mb{a}
	=
	\tn{diag}(\conj{\tb{u}}_{k-1})\mb{G}^H\mb{a} + 
	\Big[
	\frac{\partial \tb{u}_{k-1}}{\partial \mb{f}}
	\Big]^H
	\tn{diag}(\mb{f})
	\tb{G}^H\mb{a}\label{e08-grad}
	\end{align}
	for $k=1,2,...,\tn{K}$. Here, $\mb{r} = \tn{Re}\{ \mb{E}_\tn{est}\}-\mb{I}_\tn{br}$ is the residual, $\tb{A}^H$ and $\conj{\tb{A}}$ represent the Hermitian and complex conjugate of the matrix $\tb{A}$, respectively.  The recursion is initialized with $\partial \tb{u}_\tn{0}/\partial \mb{f} = \mb{0}$. 
	Brute-force evaluation of the gradient is highly computationally intensive for large scale problems, with each vector having more than a few million elements. 
	We devise a computationally efficient implementation by making use of the FFT-based structures in $\mb{G}$ and $\mb{H}$ operators.
	This algorithm extends the framework in~\cite{Kamilov.etal2016a} on small-scale 2D to large-scale 3D problems, and further demonstrates reconstruction from intensity-only as opposed to full-field measurements.
	%We provide open source code of our forward and inversion algorithm along with sample data in [REF].
	%
	\section{Results}
	\label{sec:result}
	
	We test our model on both simulations and experiments.  In our experiment, the inline holography setup uses a linearly polarized HeNe laser (632.8nm, 500:1 polarization ratio, Thorlabs HNL210L) that is collimated for illumination [Fig.~\ref{fig:setup}(b)].  
	A 4F system with a 20$\times$ objective lens (0.4NA, CFI Plan Achro) and a $200$mm tube lens is used to collect the scattered field with the Nyquist  sampling requirement satisfied. 
	A CMOS sensor (FLIR GS3-U3-123S6M-C) is used to capture the holograms.
	The object consists of polystyrene microspheres with nominal diameter $0.994~\mu \tn{m} \pm 0.021\mu$m (Thermofisher Scientific 4009A) suspended in deionized water.
	The suspension is held in a quartz-cuvette with inner dimensions $40 \times 40 \times 0.5$~mm$^3$. We are interested in localizing the individually suspended scatterers.
	A shutter speed of $5\tn{ms}$ was used and found to be sufficiently fast to capture the holograms without any motion artifacts from suspended particles.
	The illumination beam diameter is less than the width of the cuvette, while larger than the CMOS sensor area to avoid edge artifacts. 
	The front focal plane of the  objective lens was set just outside the inner wall of the cuvette for hologram recording.

	Importantly, Eq.~(\ref{e04-fmd}) requires computation of high-angle multiply scattered field propagating within the  volume; thus the internal field needs to be sampled at the Nyquist rate $\lambda / 2$. 
	In our system, the camera's pixel-size is $3.45\mu$m, and the effective lateral sampling size after magnification is $\delta x = \delta y = 172.5$nm. 
	This satisfies the sampling requirement in the medium, where the wavelength is $\lambda = 630\tn{nm} / n_\tn{water} = 473.7$nm.
	We set the voxel size along axial direction $\delta z=172.5$nm, such that the the voxels are cubic. 
	The spacing between slices is assumed to contain uniform background medium, and is set to be $5 \mu$m, approximately matching the system's  axial resolution of $\lambda / (1 - \sqrt{1-\textsf{NA}^2}) = 5.7\mu$m.
	During the computation, 2$\times$ zero-padding is used in all FFTs to avoid boundary artifacts. 
	We demonstrate large-scale inverse scattering that reconstructs  100~million voxels in a $176\times176\times500\mu \tn{m}^3$ volume.
	
	For large-scale simulation, we model the system parameters to approximately match the physical setup. 
	On such a scale, rigorous solutions like FDTD are computationally prohibitive, and sample complexity makes analytical solutions like Mie theory nontrivial. 
	We first study the effect of multiple scattering on simulated holograms using  3D SEAGLE~\cite{liu2018seagle}, which is an accurate forward model that incorporates multiple scattering, including scattering within each  particle. 
	It is based on a rigorous optimization procedure that solves the Lippmann Schwinger equation.
	We further simulate the hologram at high particle densities using our model with a sufficiently high scattering order, e.g. $\tn{K}\ge10$, such that the model converges and the simulated field closely estimates the actual. 
	In order to validate the convergence, we present an evaluation metric and show that the model converged within the first few scattering orders for all tested scenarios. 
	Improvement of our method compared to single scattering is presented quantitatively.
	
	The Boston University Shared Computing Cluster (SCC) was used for all computations. The average times of computing one iteration for the single and multiple scattering models on a $512\times512\times50$ grid
	were 58 and 257 seconds, respectively.
	All reconstructions were run for 100 iterations. 
	In what follows, we present our findings.
	
	\subsection{Effect of multiple scattering in small-scale inversion: A multislice-based approach}
	
	\begin{figure}[t]
		\centering
		\includegraphics[width=0.95\linewidth]{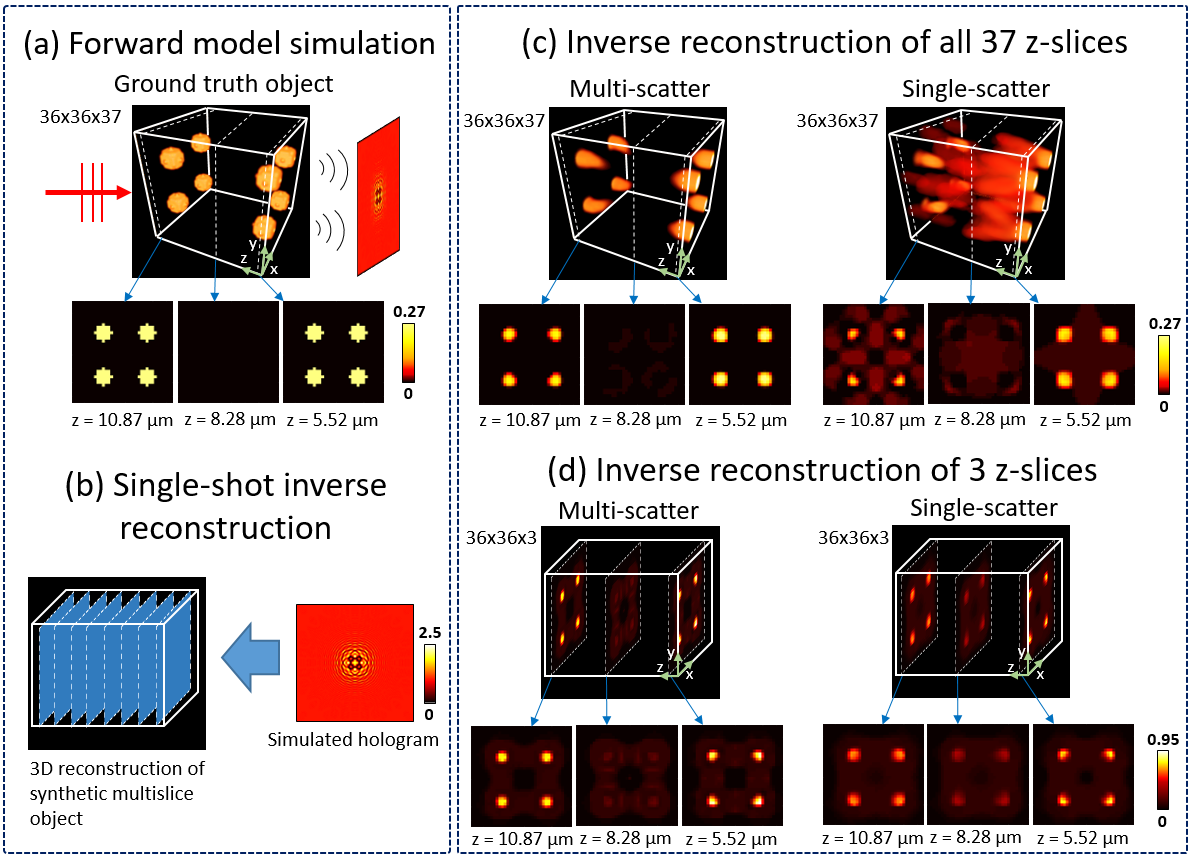}
		\caption{Small-scale multiple scattering inversion. 
		(a) An accurate 3D forward model is used to simulate the hologram. 
		(b) Multislice 3D reconstruction is performed from a single simulated measurement using our method. 
		The number of slices in the inverse reconstruction can be flexibly chosen. 
		(c) Full 3D inversion is performed by reconstructing all axial slices in the original object using our method. The multiple-scattering method outperforms the single-scattering method by providing both more accurate permittivity contrast estimation and improved optical sectioning.   
		(d) Our multislice approach enables 3D reconstruction using a much reduced number of slices while still maintaining the benefit of incorporating multiple scattering.  
		Reconstruction using only 3 slices are compared to demonstrate the improved localization capability  by our method.}
		\label{fig:ssim}
	\end{figure}
	
	It has been shown that in the presence of strong multiple scattering, the single scattering models underestimate the permittivity contrast~\cite{tsihrintzis2000higher,liu2018seagle,Ling2018}. 
	Here we validate our model on a small-scale simulation and make similar observation by showing that the underestimation is mitigated as multiple scattering is incorporated in the inversion. 
	
	The utility of our multislice-based computational approach is also demonstrated, in which the number of axial slices can be arbitrarily chosen in the inverse reconstruction.
	Effectively, we  approximate the 3D object with a fixed number of slices, such that the computation is tractable when expanding to large-scale problems.

	We simulate a volume of $44\times44\times6.4\mu$m$^3$, discretized as a $256\times256\times37$ object, containing 8 spheres in water, each with refractive index $n=$1.43 and diameter $1~\mu$m. In Fig.~\ref{fig:ssim}(a), we depict the central $6.2\times6.2\times6.4\mu$m$^3$ region of this object.
	The multiple scattering is significant in the presence of occluding geometry along the optical axis, and a refractive index contrast of $\delta n = 0.1$. It is known that occlusion causes strong axial field coupling via multiple scattering between scatterers, which is ignored by the first-order model~\cite{azimi1983distortion}. We therefore expect that incorporating multiple scattering will improve object estimation from the hologram.
	
	An inline hologram is simulated at $5 \mu m$ from the front slice using the SEAGLE.
	The hologram is then inverted using our multislice-based method incorporating $1^\tn{st}$, $2^\tn{nd}$ order scattering. The scattered intensity $|E|^2$ is included when  simulating the hologram.
	During the inversion, this term is ignored following the procedure in Sec.~\ref{sec:method}. 
	Our results indicate that even under this approximation, our model suppresses the underestimation artifacts by incorporating multiple scattering.
	
	In order to test the utility of our multislice-based approach, we perform reconstruction for two cases.
	In the first case, we reconstruct all 37 slices for the object [Fig.~\ref{fig:ssim}(c)]. The reconstruction based on the $1^\tn{st}$ order scattering underestimates the refractive indices. 
	We attribute this artifact to the strong axial coupling via multiple scattering between the occluding particles, which is ignored by the $1^\tn{st}$ order model. The underestimation is mitigated when $2^\tn{nd}$ order scattering is included in the inverse model.

	In the second case, we estimate the object using only 3 slices to perform the inverse scattering reconstruction [Fig.~\ref{fig:ssim}(d)]. 
	The reconstruction in this case approximates the 3D object comprising of 3 discrete slices. 
	We observe that our method is able to detect the 8 spheres as disks at the correct axial locations  corresponding to the centers of each particles. 
	When using only 3 slices in the reconstruction, the model has less number of slices to create the same effect at the measurement plane as the 37-slice ground truth object. In order to compensate for this, the reconstructed scattering density can be approximated as the integrated permittivity contrast along the optical axis, while still correctly localizing the particles. 
	In the $1^\tn{st}$ order result, we observe smaller contrast and worse axial sectioning. 
	The $2^\tn{nd}$ order multiple scattering  improves the contrast as well as axial sectioning, resulting in better localization capability.

	\subsection{Large-scale inversion of multiple-scattering: simulation}
	\label{sec:largescalesim}

	Next, we demonstrate the inversion of multiple scattering from single-shot measurement in large-scale. 
	For this purpose, we design a simulation which involves estimating the concentration of particles in a suspension from its inline hologram. 
	We show that our multiple scattering model improves the accuracy in estimating the particle density, particularly at larger depths.

	The simulated volume is $88\times88\times250\mu$m$^3$, discretized on a $512\times512\times50$ grid, in which disk-shaped scatterers of $1~\mu$m diameter and constant refractive index are suspended randomly in water, in varying densities.
	The disk-shape may not represent actual spheres, as would be in the real application, however it is taken as an approximation due to stringent sampling requirements for such a large volume.
	We  consider two values of the refractive index contrast $\delta n = 0.01$ and 0.19.
	For each volume, we first simulate holograms using $20^\tn{th}$ order scattering, followed by the reconstruction using $1^\tn{st}$ and $2^\tn{nd}$ order models. 
	The  particle density is estimated for each reconstructed volume  using the ImageJ 3D objects counter toolbox~\cite{Bolte2006}.
	The optimal threshold parameter used for calculating the density is determined using the ROC.
	
	\begin{figure}[t]
		\centering
		\includegraphics[width=0.9\linewidth]{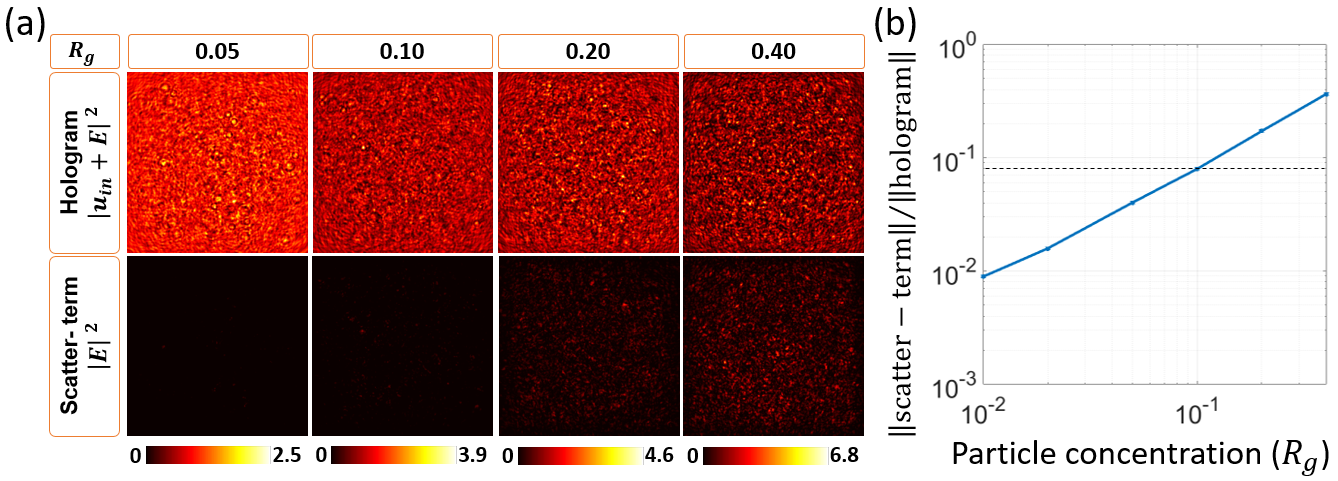}
		\caption{Effect of particle density on the scattered intensity term $|E|^2$ contribution in the hologram.
		(a) Contribution is negligible compared to the hologram for low particle densities, and becomes gradually important as the particle density increases.
		(b) The ratio of between the total intensity of the hologram and the $|E|^2$ terms for all values of $R_\tn{g}$ tested in the simulation. For $R_\tn{g} \leq 0.1$, the total intensity of the hologram is at least an order of magnitude larger than the  $|E|^2$ term.}
		\label{fig:E2}
	\end{figure}

\begin{figure}
		\centering
		\includegraphics[width=\linewidth]{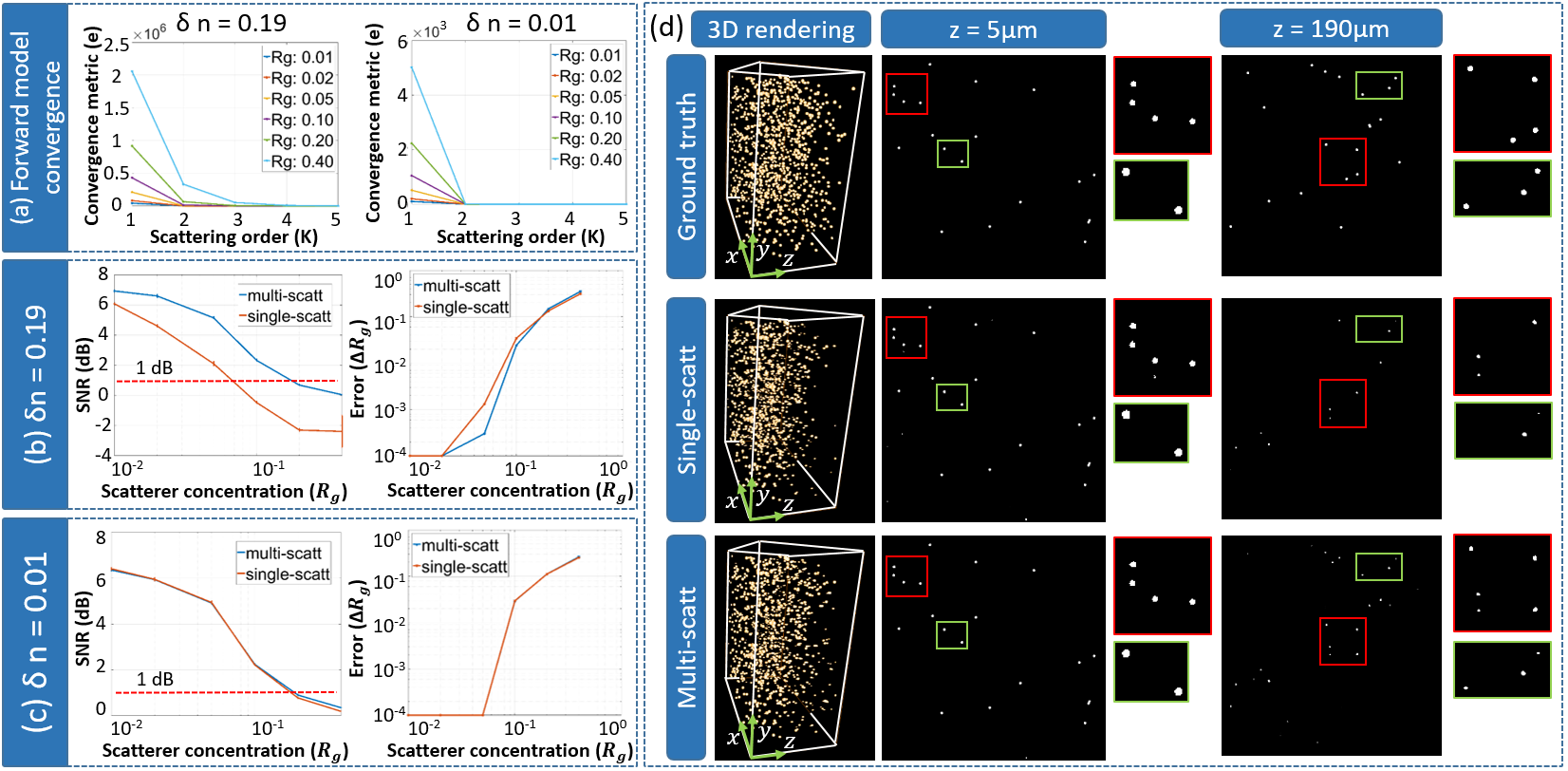}
		\caption{Validation of our multiple scattering method on large-scale simulation.
		(a) Convergence properties of the forward model are studied under varying particle densities. 
		Higher-order scattering is generally required for convergence when the object is strongly scattering. 
		In most cases studies, 2 orders of scattered field sufficiently capture the majority of the contribution.
		(b) For higher refractive index contrast ($\delta n = 0.19$), multiple-scattering performs similarly to single-scattering for low concentration ($R_\tn{g}\leq0.02$), and better than single-scattering for $0.02 < R_\tn{g}\leq 0.1$.
		Reconstruction fails for very high concentration ($R_\tn{g}>0.1$), i.e. when the SNR drops below an empirically chosen value of 1dB. 
		The error in the predicted vs the ground truth particle concentrations also shows a similar trend.
		(c) For lower contrast ($\delta n = 0.01$), multiple scattering contributions are negligible and both methods give similar performance. 
		(d) 3D rendering depicting localized particles are shown for $\delta n = 0.19$ and $R_\tn{g} = 0.1$. Both methods have similar performance for slices close to the image plane, but our multiple-scattering model performs better at increased depths.}
		\label{fig:lrg_sim}
	\end{figure}

	As a measure of particle density, we consider the geometric cross-section $R_\tn{g}$, which corresponds to the fraction of the hologram area directly occluded by the scatterers, defined as
	\begin{equation}
	R_\tn{g} = \frac{\tn{total cross-sections of all scatterers}}{\tn{area of the hologram}} \approx \frac{N_\tn{p} \pi r^2}{N_x N_y \delta x \delta y},
	\label{e09-rg}
	\end{equation}
	where $N_\tn{p}$ represents the total number of scatterers in the volume.
	This metric is valid for scattering domains which are not very thick, as in our case.
	For a collection of identical particles suspended in a homogeneous medium, the geometric and scattering cross-sections are directly related~\cite{Ishimaru1978}, in which the latter is a direct measure of the fraction of the incident light scattered by an object. 
%	This interpretation is useful because the scattering cross-section is 
	For higher values of $R_\tn{g}$, we thus expect greater contributions of multiple scattering. 
	From the signal processing perspective, $R_\tn{g}$ also  measures  the sparsity of the problem as it approximates the ratio between the number of nonzero unknowns to the number of measurements.
	The values of $R_\tn{g}$ tested are 0.01, 0.02, 0.05, 0.1, 0.2 and 0.4, corresponding to $N_p = 100,~200,~500,~1000,~2000$, and $4000$, respectively. 
	We simulate five random object volumes for each value of $R_\tn{g}$ and refractive index contrast, and report the mean statistics of the reconstructions in Fig.~\ref{fig:lrg_sim}.

	In Sec.~\ref{Forward model}, we assumed that the intensity of the scattered field $|E|^2$ is negligible in the forward model. 
	In this study, this assumption holds true when $R_\tn{g}\le0.1$, where the contribution of $|E|^2$ is at least an order of magnitude smaller than the total intensity of the hologram [Fig.~\ref{fig:E2}].
	For higher particle density, $|E|^2$ becomes increasingly significant, which leads to greater model error.

For the series expansion approach used in our model, it is important to evaluate its convergence.  
	In Fig.~\ref{fig:lrg_sim}(a), we present the convergence properties of the forward model under our experimental conditions.
	While in general higher-order terms are required for convergence under  stronger scattering, the $2^{\tn{nd}}$ order scattering is sufficient for most of the cases studied. 
	Our convergence metric $e$ is defined by the residual error of the field within the 3D volume~\cite{vandenBerg.Kleinman1997,Mudry.etal2012,liu2018seagle}, as
	\begin{equation}
	e = \|\mb{A}\mb{u}_\tn{K} - \mb{u}_\tn{0}\| \label{conv_metric},
	\end{equation}
	where $\mb{A} = \mb{I} - \mb{G}\tn{diag}(\mb{f})$. 
	This convergence metric essentially measures the self-consistency of the total internal field~\cite{liu2018seagle}. For $\tn{K}$-order scattering, it computes the norm of the residual contribution from ($\tn{K}+1$)-order scattering, which must approach zero in the case of convergence. 
	
	\begin{figure}
    	\centering
    	\includegraphics[width=\linewidth]{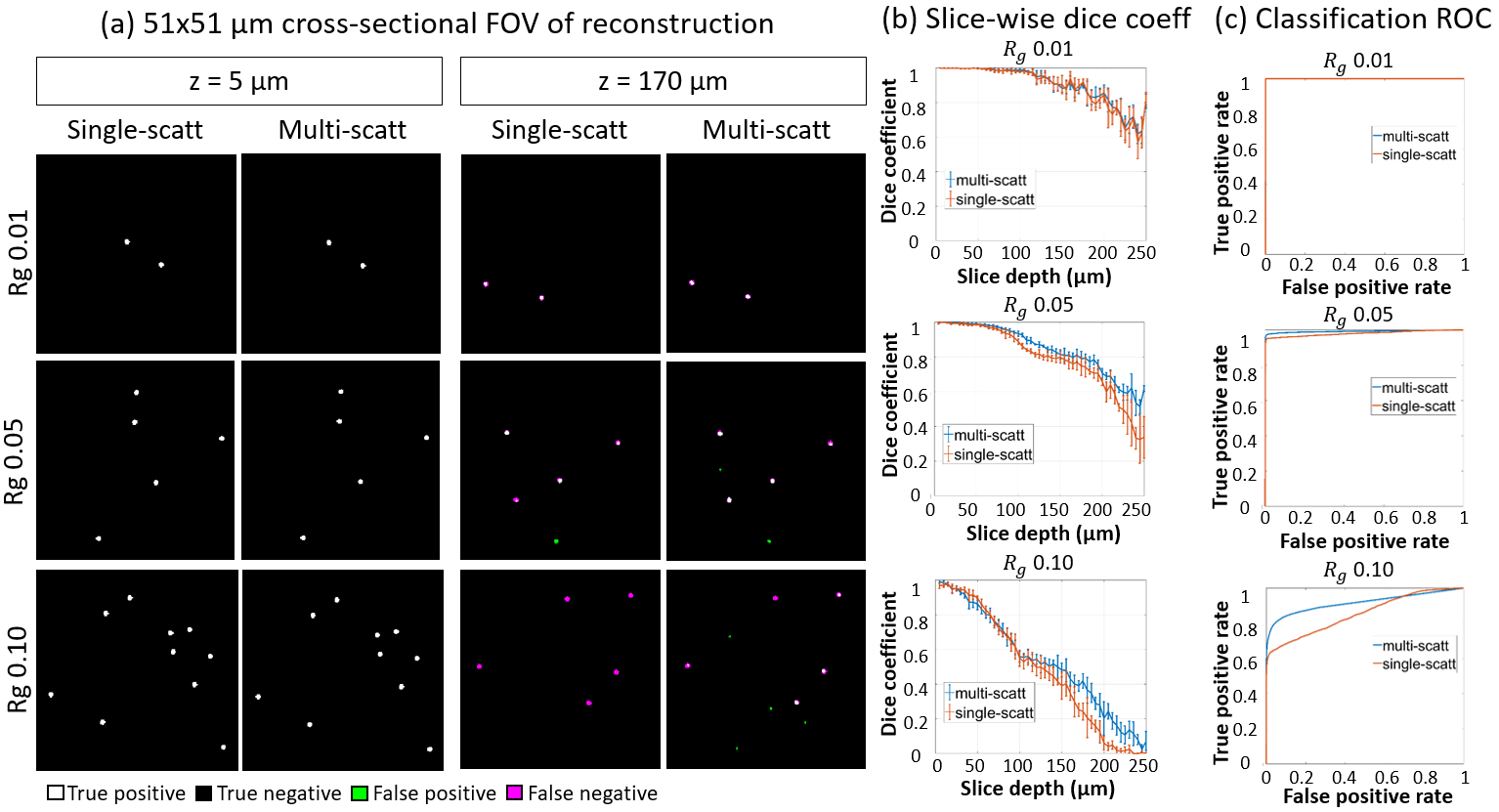}
    	\caption{Reconstruction performance as a function of depth.
    	(a) Segmentation maps of reconstructed slices (zoomed-in 51x51$\mu$m$^2$ regions) at different depths (true positive: white, true negative: black, false positive: green, false negative: pink). For object slices close to the hologram, both multiple and single scattering methods provide high accuracy. 
    	At larger depths, the accuracy deteriorates for both methods. 
    	Our multiple scattering method performs notably better at larger depths for higher particle densities.
    	(b) The slice-wise dice coefficient plotted as a function of slice depth also indicates that the multiple scattering model provides improved segmentation accuracy, especially at greater depth.
    	(c) The particle localization accuracy is quantified using the ROC curve.  
    	The curves corresponding to the multiple scattering solutions consistently have larger areas underneath, indicating better localization accuracy as compared to the single scattering method in all cases studied. 
     	}
    	\label{fig:ROC}
	\end{figure}

	Next, we evaluate the reconstruction accuracy by measuring the signal to noise ratio (SNR)
	\begin{equation}
	    \mathrm{SNR} = 20 \log_{10}\Bigg( \frac{\|\mb{f}_\tn{true}\|}{\|\mb{f}_\tn{true}-\hat{\mb{f}}\|}
	    \Bigg),
	\end{equation}
	where $\mb{f}_\tn{true}$ and $\hat{\mb{f}}$ are the true and estimated objects, respectively.
	For higher index contrast ($\delta n=0.19$), our multiple-scattering model performs consistently better than the single-scattering model for all densities tested [Fig.~\ref{fig:lrg_sim}(b,c)]. 
	Generally, the reconstruction performance from both methods drop as the density  increases. 
	We attribute this to stronger higher-order scattering and decrease in the object sparsity.  
	The stronger scattering introduces higher order nonlinearity through Eq.~\eqref{e04-fmd}, making the problem harder to invert. 
	The decrease in object sparsity leads to an effective smaller measurement-to-unknown ratio, further worsen the ill-posedness of the problem.

	For higher contrast ($\delta n = 0.19$), at low particle density ($R_\tn{g}\leq0.02$), single and multiple scattering methods perform similarly. 
	This is expected since multiple scatterings are weak due to the small scattering cross section.
	For $0.02 < R_\tn{g}\leq~0.1$, our method outperforms the single scattering method, providing a better estimate of the actual particle density. 
	For $R_\tn{g}>0.1$, the SNR drops below 1dB, and we empirically consider the reconstruction has failed [Fig.~\ref{fig:lrg_sim}(b)].

	For lower index contrast ($\delta n=0.01$), both multiple and single scattering methods perform almost identically  for all densities tested, which indicates that the contribution from multiple scattering is negligible for low refractive index contrast [Fig.~\ref{fig:lrg_sim}(c)]. 3D renderings and cross-sectional reconstructions at different depths are depicted in Fig.~\ref{fig:lrg_sim}(d).

	The depth-dependent performance is highlighted in Fig.~\ref{fig:ROC}(a,b).
	Close to the hologram plane ($z=5\mu$m), single and multiple scattering reconstructions are similar, and match the ground truth. 
	At larger depth ($z=190\mu$m), the single-scattering reconstruction degrades and results in a large number of missing particles [Fig.~\ref{fig:ROC}(a)]. 
	Our multiple scattering model improves the localization at larger depths. 
	By treating particle localization as a binary classification problem, we use the ROC curve to determine the optimal segmentation threshold when quantifying the voxels reconstructed by each method~\cite{Chen.etal2015a}.
	This allows us to evaluate the localization accuracy slice-by-slice, whose statistics are accumulated by 5 different object volumes for each particle density. The statistic we use for comparison is the dice coefficient which is used to gauge the similarity of two samples and is defined as
	\begin{equation}
	    \mathrm{D} = \frac{2\sum_{i=1}^{N} p_i g_i}{\sum_{i=1}^{N} {p_i}^2+\sum_{i=1}^{N} {g_i}^2},
	\end{equation}
	where $p_i$ and $g_i$ each represents a voxel from the predicted and ground truth binary segmentation volumes respectively and $i$  indexes the voxels of each 3D volume.
	%%%% Define dice coefficient
	%%%%
	Consistent with the visual inspection in Fig.~\ref{fig:ROC}(a), the dice coefficient clearly indicates improvement at larger depth using our multiple scattering model [Fig.~\ref{fig:ROC}(b)].
	In addition, the area under each ROC provides a direct measure of the algorithm's overall classification performance.
	Our results indicate that the multiple scattering model consistently outperforms the single scattering method [Fig.~\ref{fig:ROC}(c)].
	
	%As the depth increases, the single-scattering reconstruction degrades rapidly and . 
	%The degradation of axial and lateral resolutions is known for the single scattering case, and can be attributed to the decrease in NA with increase in depth. 
	%For the case of multiple scattering, we account for high-NA inter-slice propagations within the object. 
	%We suspect that this may have an effect similar to increasing the effective NA, which becomes more prominent in slices deeper within the 3D object.
	% (d)

	%From the figure, it can be observed that convergence is achieved in all cases. 
	%For low contrast even $2^{nd}$ order scattering suffices.
	%For higher contrast, the order required for convergence increases with particle concentration. For the same particle density, the sample with lower refractive index demonstrates better convergence.
	%
	
	\begin{figure}
		\centering
		\includegraphics[width=.75\linewidth]{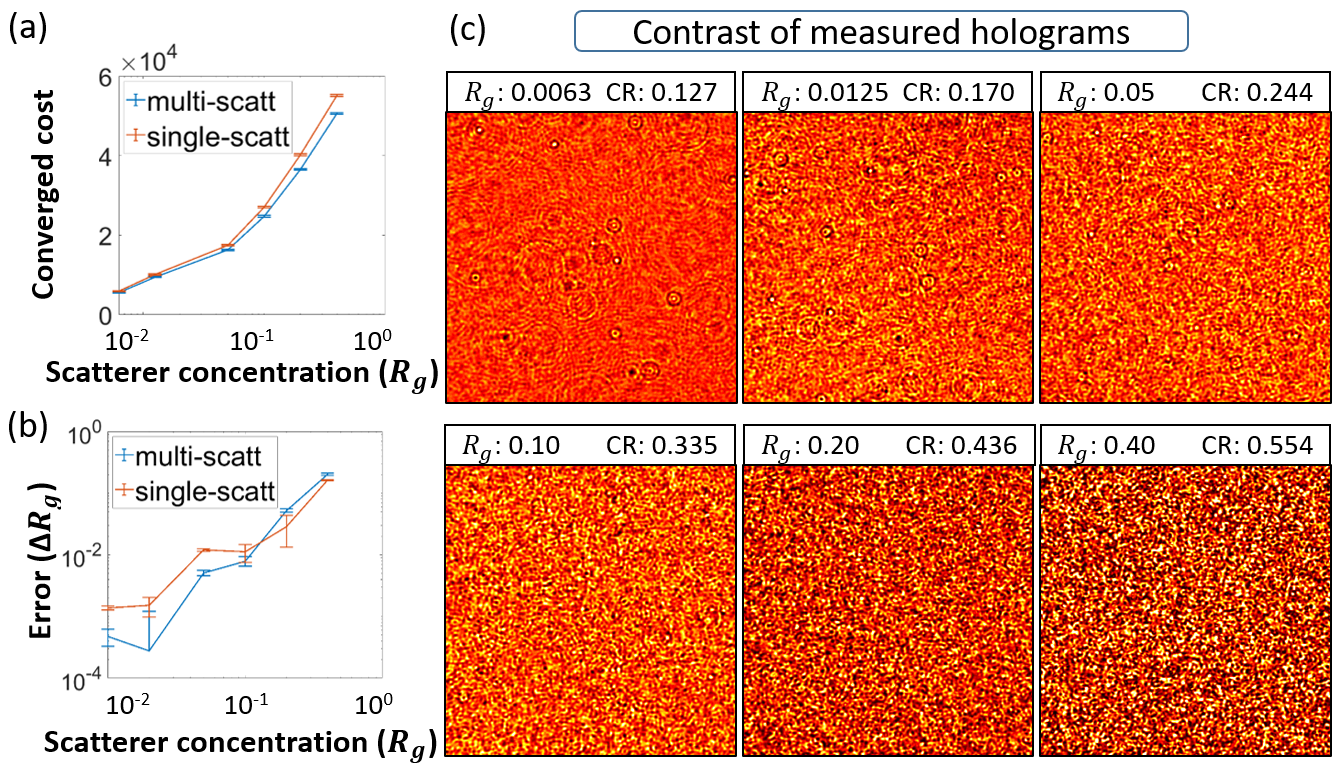}
		\caption{Experimental validation of our method in large-scale. (a) The multiple-scattering model converges to a lower cost than the single-scattering model for all concentrations indicating better fit to the cost function. 
		(b) The reconstructed particle density follows a trend similar to the simulation where multiple-scattering performs better than the single scattering method for $R_\tn{g} \leq 0.1$; both methods fail for $R_\tn{g}>0.1$. 
		(c) As $R_\tn{g}$ increases, the hologram gradually resembles speckle patterns, as quantified by the contrast ratio~(CR). }
		\label{fig:exp1}
	\end{figure}	
	
	\begin{figure*}
		\centering
		\includegraphics[width=\linewidth]{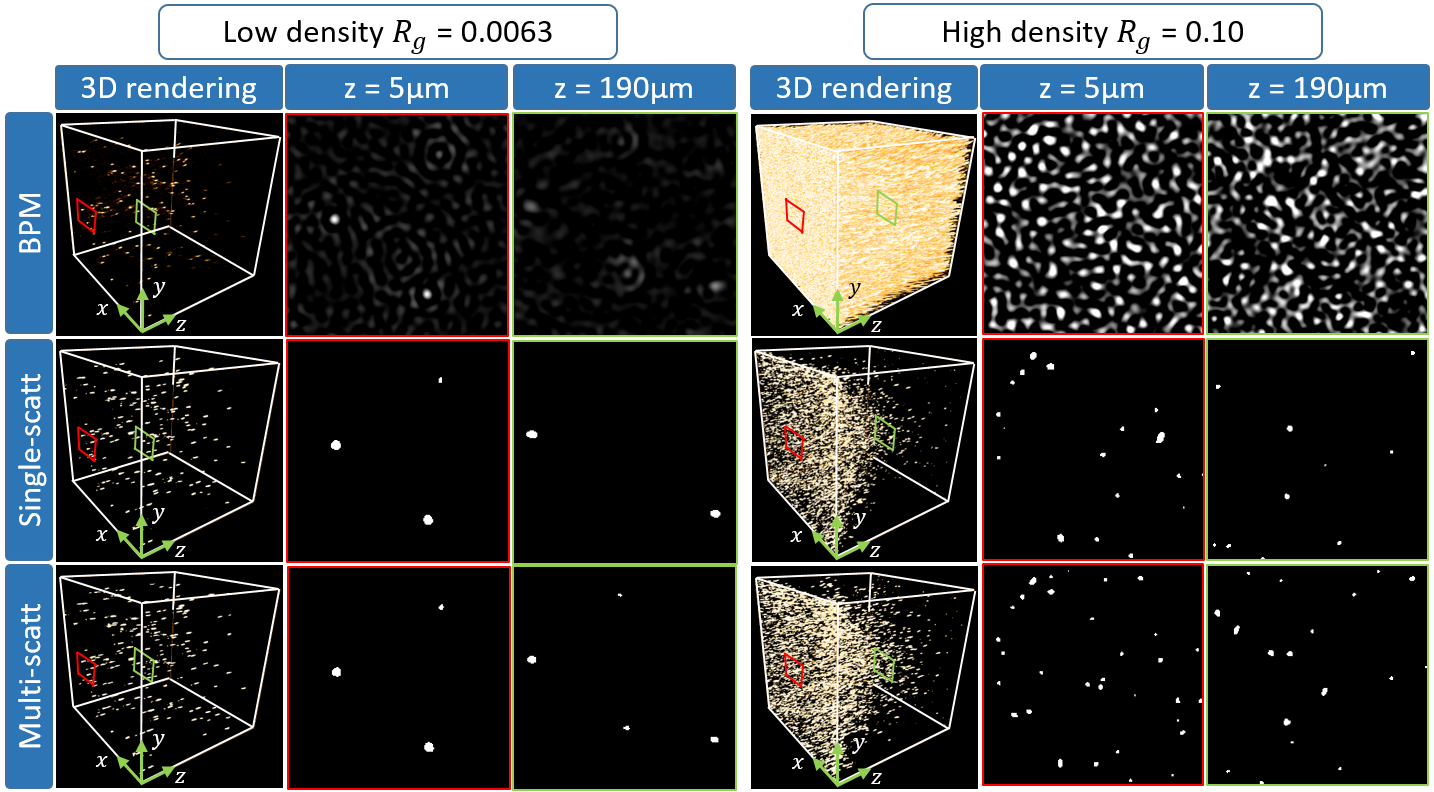}
		\caption{3D visualization of the localized particles under different concentrations from our experiment and their $200 \times 200 $ lateral cross sections at different depths.
		For low density, both multiple and single scattering methods perform similarly. For high density,  the underestimation of particles from the single scattering method is clearly visible, especially at increased depth. 
		Our multiple scattering model mitigates the underestimation as it accounts for the inter-coupling between particles whose strength increases as the depth.
		The traditional BPM is effective for low density, but completely fails for high density and the reconstruction resembles speckles throughout the volume.}
		\label{fig:exp2}
	\end{figure*}
	
	\subsection{Large-scale experimental validation}
	% a lot of particles experiment test
	Finally, we demonstrate our method on a set of large-scale experiments.
	We reconstruct over 100 million object voxels ($1024\times1024\times100$) from each 1-megapixel hologram.
	Our multiple scattering model significantly improves the 3D localization accuracy as compared to the BPM and single scattering methods.
	Notably, {\it our experimental results closely match the simulation}.
	
	We prepare polystyrene microsphere suspensions, ranging from dense to sparse concentrations via successive dilution, with corresponding $R_\tn{g}$ values of 0.4, 0.2, 0.1, 0.05, 0.0125, 0.0063.
	Five holograms are recorded at each concentration, and then used for reconstructions and density estimation. 
	Background subtraction is performed on each hologram as a preprocessing step to remove static artifacts.
	The inversion is performed using our method with second-order multiple scattering ($\tn{K}=2$), the single scattering method, and BPM. 
	
	First, we evaluate the results based on the optimization convergence cost [Fig.~\ref{fig:exp1}(a)]. 
	The multiple scattering method converges to a lower value than single scattering for all densities, indicating  better fit to the cost function $\mathcal{C}(\mb{f})$.
	The cost increases for both methods with $R_\tn{g}$, depicting degradation of reconstruction with increase in particle density.

	Next, we assess the estimated particle density.
	Our multiple scattering model consistently performs better than the single scattering model for $R_\tn{g}\leq0.1$ [Fig.~\ref{fig:exp1}(b)]. 
	For $R_\tn{g} > 0.1$, reconstruction fails for both methods as also found in our simulation. 
	Hence, we use  $R_\tn{g}=0.1$ as an empirical performance bound of our method for this application.
	
	Evidently, the recorded holograms gradually resemble speckle patterns as the particle density increases [Fig.~\ref{fig:exp1}(c)]. 
	We quantify the hologram's contrast ratio (CR) at each density, which can be used as an alternative metric. The CR is calculated as the ratio of the standard deviation to the mean~\cite{Ishimaru1978}.
	At the critical $R_\tn{g}=0.1$ concentration, the CR is around 0.335.

	Finally, we closely examine the 3D reconstructions for $R_\tn{g}=0.0063$ and $R_\tn{g}=0.1$ [Fig.~\ref{fig:exp2}].
	For the low density case, single and multiple scattering methods perform similarly due to weak multiple scattering.
	For the high density case where multiple scattering becomes significant, our method outperforms the single scattering model.
	Particle localization degrades with increased depth;
	our multiple-scattering method provides a more uniform estimation and better localization at increased depth, matching our observations in the simulation. 
	While BPM is able to reconstruct individual particles at the low density, it completely fails for high density, and resembles speckles extended across the object volume.
	%

	%\section{Discussion}
	%% talk about convergence of experimental 
	%While the forward model was shown to converge for the simulated samples, it is not always the case. In Fig.~\ref{fig:conv_exp} we show convergence plots for the reconstructed objects (multiple-scattering method) from the physical experiment. It is observed that for high density, our model does not converge. This might also shed some light into why the reconstructions fail at very high densoty. However, in-depth study would be required to comment on this point conclusively. One extension of our work may be to incorporate the convergent Born series \cite{Osnabrugge2016}, in which the Green's function is modified to mathematically guarantee convergence even for strongly scattering samples. 
	%
	%\begin{figure}[t]
	%	\centering
	%	\includegraphics[width=\linewidth]{figure_8}
	%	\caption{Our forward model was used to simulate the $10^\tn{10}$-order internal field within the object fro the reconstructed objects from the physical experiment. It was observed that the field convergence was fast for $R_\tn{g}<0.1$ and slow when  $R_\tn{g}=0.1$. For $R_\tn{g}<0.1$, the field did not converge, and subsequent multiply-scattered fields kept increasing in strength.}
	%	\label{fig:conv_exp}
	%\end{figure}
	
	\section{Conclusion}
	
	We have presented a new computational framework for utilizing multiple scattering in in-line holography for large-scale 3D particle localization. 
	Our model recursively computes both forward and backward multiple scattering in a computationally efficient manner. 
	Both simulations and experiments demonstrate the significance of modeling multiple scattering in alleviating depth-dependent artifacts and improving the 3D localization accuracy compared to traditional methods.   
	Our method may opens up new opportunities for large-scale imaging applications utilizing multiple scattering. 
	
	Our model is currently limited by the convergence regime of the classical Born series expansion, preventing it to be applied to particle density higher than 0.1 geometric cross-section. 
	Recent work on convergent Born series expansion~\cite{Osnabrugge2016} provides a promising avenue to extend our model to higher scattering scenarios.  
	
	The novel multislice structure proposed in our model provides a flexible framework for trading computational cost for model accuracy.
	Still, higher order scattering calculation necessitates longer computational times, which is less appealing for applications requiring real-time reconstructions. 
	To facilitate rapid volumetric estimation without sacrificing accuracy, recent machine learning based inverse scattering approaches~\cite{sun2018efficient,li2018imaging,li2018deep} may be explored in the future. 

\section*{Disclosures}

There are no conflicts of interest related to this article.
	
\section*{Funding}

National Science Foundation (NSF) (1813848, 1813910).

\section*{Acknowledgments}
The authors thank Anne Sentenac, Alex Matlock, and Benedict Diederich for helpful discussions regarding multiple scattering,
Yujia Xue for discussions on the simulations, and Quentin Lin for help on the experiment.

%%%%%%%%%% If using BibTeX:

\bibliographystyle{spiejour}

\end{spacing}
\end{document}